\documentclass[a4paper,fleqn,usenatbib]{mnras}

\usepackage[T1]{fontenc}
\usepackage{ae,aecompl}

\usepackage{graphicx}	
\usepackage{amsmath}	
\usepackage[psamsfonts]{amssymb}	
\usepackage{lmodern}   

\title[The Progenitor Model of SN 1987A]{Progenitor Model of SN 1987A Based on the Slow Merger Scenario}

\author[Urushibata et al.]{
Takaki Urushibata,$^{1}$\thanks{E-mail: urushibata@astron.s.u-tokyo.ac.jp}
Koh Takahashi,$^{2}$
Hideyuki Umeda$^{1}$
and Takashi Yoshida$^{1}$
\\
$^{1}$Department of Astronomy, Graduate School of Science, University of Tokyo Hongo 7-3-1, Bunkyo-ku Tokyo 113-0033, Japan  \\
$^{2}$Argelander-Institut f\"{u}r Astronomie, Universit\"{a}t Bonn D-53121 Bonn, Germany   \\
}

\date{Accepted XXX. Received YYY; in original form ZZZ}

\pubyear{2017}

\begin{document}
\label{firstpage}
\pagerange{\pageref{firstpage}--\pageref{lastpage}}
\maketitle

\begin{abstract}
Even after elaborate investigations for 30 years, we still do not know well how the progenitor of SN 1987A has evolved. 
To explain unusual red-to-blue evolution, previous studies 
suggest that in a red giant stage either the increase of surface He abundance 
or the envelope mass was necessary. It is usually supposed that the He enhancement is
caused by the rotational mixing, and the mass increase is by a binary merger. 
Thus, we have investigated these scenarios thoroughly. The obtained findings are that 
rotating single star models do not satisfy all the observational constraints and 
that the enhancement of envelope mass alone does not explain observations.
Here, we consider a slow merger scenario in which both the He abundance and 
the envelope mass enhancements are expected to occur. We indeed show that 
most observational constraints such as the red-to-blue evolution, lifetime, total mass, 
position in the HR diagram at collapse, and the chemical anomalies are well reproduced by the merger model of 14 and 9 M$_{\odot}$ stars.
We also discuss the effects of the added envelope spin in the merger scenarios. 

\end{abstract}




\section{Introduction}

SN 1987A, which occurred in the Large Magellanic Cloud (LMC),
is the most well studied supernova (SN) that has ever observed.
Thanks to its proximity to the earth,
numbers of observations
including the precise light curves and spectroscopies,
the finding of the progenitor in pre-explosion images,
and the detection of neutrinos
have been obtained to manifest that
core collapse of a massive star indeed triggers the supernova
\citep{gilmozzi+87,hirata+87,woosley88,suntzeff&bouchet90}.
For theory of stellar evolution, however,
the observation casts a perplexing question:
the progenitor was a blue-supergiant (BSG) and once had been a red-supergiant (RSG).

The theory of stellar evolution predicts that
a massive star with an initial mass of $8$--$20 {\rm M}_\odot$
ends its life as an RSG.
However, several observations indicate the compactness of the envelope of SN 1987A.
Firstly, the light curve does not have the plateau phase,
which characterizes type II-P SNe having inflated RSG envelopes.
Secondly, neutrinos are detected just three hours before the shock break out,
which will take a few days to propagate through the RSG envelope.
Finally, the progenitor found in the pre-explosion images, Sk -69$^\circ$202,
has a hot blue surface of log $T_{\rm eff} \simeq 4.2$.
Furthermore, it is considered that once the progenitor has been a RSG.
Surrounding the supernova remnant, three ring-like nebulae with a radius of
$\sim 300$ light-days have been discovered \citep{wampler&richichi89}.
Since the high He abundance of {\rm He/H} $= 0.17 \pm 0.06$ and the CNO abundance ratios of {\rm N/O} $ = 1.5 \pm 0.7$ and {\rm N/C} $ = 5.0 \pm 2.0$
are compatible to hydrogen burning products \citep{lundqvist&fransson96,mattila+11},
and since the expansion velocity of $\sim 10$ km s$^{-1}$ in the nearest ring from the explosion position is
comparable to the wind velocity of RSGs,
these nebulae are considered to consist of ejecta from an RSG.
This, in turn, indicates Sk -69$^\circ$202 was a RSG
at least $\sim 2 \times 10^4$ yr ago \citep{crotts&heathcote01}.

To begin with, we have tried to reproduce the envelope evolution
by applying some scenarios proposed so far (\citealt{urushibata+17a}; Urushibata et al. 2017\natexlab{b} in prep).
We adopt two kinds of scenarios based on a single star or a binary model.
In a single star model, a red-to-blue evolution in \citet{saio+88b} is supposed to be triggered by the efficient internal mixing
that dredges the helium core material up to the convective envelope.
This mixing simultaneously decreases the core mass to the total mass ratio and
increases the helium mass fraction in the envelope, both of which helps the deflation \citep{saio+88a}.
On the other hand,
\citet{podsiadlowski+92} showed that envelope mass enhancement
due to mass accretion or stellar merger in a binary system helps to deflate the RSG envelope.



To investigate a role of the effects, we first tried to increase the helium abundance in the envelope due to rotational mixing using our stellar evolution code with rotational effects \citep{takahashi+14}.
We found that rotation mixing alone is almost impossible to satisfy the observational constraints including the timescale of the red-to-blue evolution.
Next we added matter from the surface, investigating importance of envelope mass increase.
The composition of the matter is set to be the surface composition of the added star.
Here we set mass of the matter, duration of the accretion, and evolutionary phase of the star when we start to add the matter as parameters.
We found that the transition to a BSG is relatively easy if we add the mass just after the primary becomes a RSG. 
Unfortunately this situation contradicts with the observational constraint that the progenitor was a RSG at least $\sim 2 \times 10^4$ yr ago. 
This condition indicates that the mass enhancement should take place during the CO core formation with fully the convective envelope.
In this case the transition to the BSG is much difficult. 
We added 15 ${\rm M}_{\odot}$ as maximum to the envelope, but a RSG never moves bluer. 
Therefore we concluded that the satisfactory transition cannot occur in a simple binary model.



Thus in this Letter, we move on to construct a progenitor model based on the slow merger scenario \citep{ivanova+02}, expecting both of the effects of the strong internal mixing and the envelope mass enhancement to take place together.
Moreover, highly anisotropic mass ejection to account for the triple-ring nebulae, which will be unlikely to happen in single models
except for a fast rotating model \citep{podsiadlowski+92}, can be explained by this scenario \citep{morris&podsiadlowski07}.

The picture is as follows.
Initially two stars form a binary system with a separation of $\sim$ 1000 R$_\odot$.
The binary forms a common envelope after or during the primary's helium burning phase.
Inside the common envelope, the primary's core and the secondary rapidly get close
by transporting the orbital angular momentum to the envelope.
At some point, the secondary's Roche radius becomes smaller than its stellar radius,
and the secondary starts to melt into the envelope.
A hydrodynamical simulation by \citet{ivanova+02} has shown that
a streamline of the melting material can penetrate the $\mu$-barrier at the hydrogen/helium boundary.
The penetration results in rapid hydrogen burning, 
by which turbulence is powered and mixes the region homogeneously.
After the merger, the secondary-accreted-primary star restarts the evolution,
changing its structure to find a new thermally equilibrated state.

In the next section, we give a description of our one-dimensional approach,
in which matter accretion, internal chemical mixing, and acceleration of the envelope spin
are taken into account.
Then we show the impact of slow merger on the primary's evolutionary properties.
Since our model has numbers of controlling parameters
and accordingly numerous models with different parameters have been calculated,
a representative model,
which provides a satisfactory explanation for observational constraints of the SN 1987A progenitor,
is selected for detail discussion for brevity.
Finally we give our conclusions.

During preparation of this letter, a similar attempt to model the evolution of
Sk -69$^\circ$202 has been presented by \citet{menon&heger17}.
Apart from the line of reasoning why we consider the slow merger scenario,
the acceleration of the envelope spin is what we have newly considered.
Although only our work have reproduced both of the observational constraints of
the surface N/C ratio and the timing of the blueward evolution of $2 \times 10^4$ yr before the collapse,
the red-to-blue evolution have been obtained in both works.


\section{Method}

Simulations have been done with the stellar evolution code described in \citet{takahashi+14}.
We consider a binary of masses of $M_1$ and $M_2$.
First, the primary evolution is solved as a single stellar evolution until its collapse.
Then, taking a model $2 \times 10^4$ yr before the collapse as an initial condition,
the slow merger process is solved by a one-dimensional approach.
In the merger calculation, the mass-averaged chemical composition of the secondary is used,
which is calculated by another evolution calculation.
The evolution of the merger product is continuously solved until the final collapse.

Solar scaled metallicity \citep{asplund+09} with a reduction factor of 0.355
is adopted to account for the LMC iron abundance.
The prescription of \citet{vink+00, vink+01} is used for wind mass loss of
hotter models with log $T_{\rm eff}$ [K] $> 3.9$,
while the rate by \citet{nieuwenhuijzen&dejager90}
with a metallicity reduction factor of ($Z$/$Z_\odot$)$^{0.64}$ is used for cooler models.
For opacity, tabulated data by {\it OPAL} project \citep{iglesias&rogers96}
together with the conductive opacity by \citet{potekhin+06} and the 
molecular opacity by \citet{ferguson+05} are used.
The code treats the effect of stellar rotation.
With parameters of the ratio of the diffusion coefficient to the turbulent viscosity $f_{\rm c}$ = 1/36 and the sensitivity of the rotationally induced mixing to $\mu$-gradients $f_{\mu} = 0.1$,
several mechanisms of
the meridional circulation,
the Solberg-H\o iland and Goldreich-Schubert-Fricke instabilities,
the secular and dynamical share instabilities,
and the Tayler-Spruit dynamo
are accounted for the angular momentum transfer and the rotational mixing \citep{heger+00,heger+05}.

\subsection{One dimensional approach}

We take into account the spinning-up of the stellar envelope
due to angular momentum transportation from the orbit.
With the timescale of the spinning-up, $\tau_{\rm spin}$,
the rotation of the primary's convective envelope is accelerated by
\begin{eqnarray}
	\frac{ \partial j\bigl(M\bigr)}{ \partial t } =
	\frac{ j\bigl(M\bigr) }{ \tau_{\rm spin} } \frac{ J_{\rm add} }{ J_{\rm env} },
\end{eqnarray}
where $j\bigl(M\bigr)$ is the specific angular momentum at the mass coordinate $M$,
and $J_{\rm env} \equiv$ $\int$$_{\rm envelope} j {\rm d}M$ is the total angular momentum
confined in the envelope.
The acceleration is switched-off after $\tau_{\rm spin}$ passes.
Through this process, the angular momentum of $J_{\rm add}$ in total is added to the envelope.
When the surface rotation of the primary approaches the critical velocity,
a highly efficient stellar mass loss is triggered
\citep[$\Omega \Gamma$-limit;][]{langer98,maeder&meynet00}.
The enhancement of the mass loss rate is treated as
\begin{eqnarray}
	\dot{M} = -{\rm min}\Bigl[
		|\dot{M}(v_{\rm rot}=0)| \times \Bigl( 1-\frac{v_{\rm rot}}{v_{\rm crit}} \Bigl)^{0.43},
		0.1 \frac{M}{\tau_{\rm KH}}
	\Bigl], \label{og-limit}
\end{eqnarray}
where $v_{\rm rot}$ and $v_{\rm crit} \equiv \sqrt{ GM\bigl(1-L/L_{\rm Edd}\bigr)/R }$ are
the rotation velocity and the critical rotation velocity at the surface of the star \citep{yoon+10,yoon+12}.
The mass loss ejects the spin angular momentum as well.
Currently spherical mass loss is assumed to take place so that
the ejected material has a specific angular momentum of the stellar surface.
Possible over reduction of the angular momentum due to disk-like mass ejection
may affect the evolution of the spinning-up envelope.
We shortly discuss this point later.

We do not take the effect of frictional heating into account in this work.
\cite{morris06} found that quantity of deposited energy due to the heating could affect on geometry of the ejecta, 
and that mass ejection in the equatorial plane was suppressed if the deposited energy was less than one-third of the binding energy of the envelope, meaning the condition for creating the complexed nebula.
In a current stellar evolution code, it is difficult to handle this phenomena consistently.

The next effect we consider is the melting of the secondary star with the envelope.
The three-dimensional geography resulting from the interaction between
a thin stream line of the melting material and the primary's core
is expressed as a one-dimensional chemically homogenized region.
Three parameters are set:
$R_{\rm out}$, the end point of the turbulent mixing region;
$M_{\rm in}$, the reaching point of the stream;
and $\tau_{\rm melt}$, the timescale of the melting.
Equivalently, a mass coordinate at $R_{\rm out}$ is determined as $M_{\rm out}$ in each time step.
During the melting time of $\tau_{\rm melt}$,
a mass of $M_2$ in total is inserted to the base of the envelope, and
envelope chemical composition between $M_{\rm in}$ and $R_{\rm out}$ are modified.
With the time step of $\Delta t$, the new chemical compositions are determined as
\begin{eqnarray}
	X_i = \frac{ \bar{X}_{i, b}( M_{\rm out}-M_{\rm in} ) + X_{2, i} \Delta M }
	           { M_{\rm out} + \Delta M - M_{\rm in} },
\end{eqnarray}
where $X_i$ is the mass fraction of the $i$-th isotope,
$\bar{X}_{i, b} \equiv \int_{M_{\rm in}}^{M_{\rm out}} X_{i, b} {\rm d} M$
is the mean mass fraction at the previous step,
$X_{2, i}$ is the composition of the secondary,
and $\Delta M \equiv \bigl(\Delta t / \tau_{\rm melt}\bigr)M_2$
is the accreted mass during one time step.

As well as the internal chemical profile,
profiles of other conserved quantities of the angular momentum and the entropy
should be determined to model the accretion.
We currently make these profiles just by stretching profiles at the previous step in mass coordinate.
The orbital and secondary's spin angular momenta are thus neglected.
Also the complex energetics of the melting,
such as releasing the orbital kinetic energy, are significantly simplified.
The effects of these simplifications will be investigated in future.


\subsection{Parameter ranges}

The secondary mass is selected from $M_2 \in [1.2$--$10]$ M$_\odot$
in order to have a radiative envelope during the main sequence phase.
A constraint for the ejecta mass of SN 1987A of $M_{\rm ej} = 18 \pm 1.5$ M$_\odot$
is obtained through analyses of
the bolometric light curve and the spectral evolution of $H_\alpha$ lines \citep{utrobin07}.
Adding the mass of the neutron star, the total mass of the system at the explosion is estimated to be
$M_{\rm tot} \sim 19.4 \pm 1.5$ M$_\odot$.
To match with this constraint, the range of the primary mass is set as $M_1 \in [14$--$18]$ M$_\odot$.

As for timescales of $\tau_{\rm spin}$ and $\tau_{\rm melt}$, a typical value of 100 yr is set (see e.g. \citealt{ivanova+02}).
A fiducial value of the additional angular momentum may be estimated
by considering revolving two point masses of $m_1$ and $m_2$ with a separation of $a$.
This gives $J_{\rm add}$ $\sim$ $3.2 \times 10^{54}$ erg s
with $m_1 = 15$ M$_\odot$, $m_2 = 5$ M$_\odot$, and $a = 1000$ R$_\odot$.
Thus we test several values for $J_{\rm lost}$ from $1\times10^{53}$ erg s up to $3\times10^{54}$ erg s.
For $M_{\rm in}$, three choices of
($M_{\rm He} - 0.2$, $M_{\rm He}$, $M_{\rm He} + 0.2$) M$_{\odot}$ are tested,
where $M_{\rm He}$ is the mass of the helium core,
which is determined by the hydrogen mass fraction being 0.01.
For log $R_{\rm out}$/R$_\odot$, (1.25, 1.75, 2.25, 2.75) are tested.

\section{Result}

\subsection{A model of SN 1987A progenitor}

\begin{figure}
 \begin{center}
    \includegraphics[height=0.45\textwidth, angle=270]{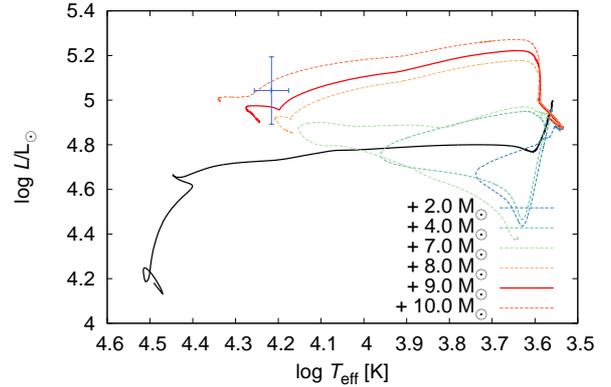}
    \caption{HR diagram.
    The observed position of Sk -69$^\circ$202 is shown by the cross \citep{arnett+89}.
    The values in the legends indicate the $M_2$ of models,
    while other parameters are same as the satisfactory model.}
    \label{fig-hr}
  \end{center}
\end{figure}

The surface evolution of a model with 
($M_1$, $M_2$, $M_{\rm in}$, log $R_{\rm out}$/${\rm R}_{\odot}$, $J_{\rm add}$) =
(14 M$_\odot$, 9 M$_\odot$, 4.6 M$_\odot$, 2.75, $3 \times 10^{53}$ erg s)
is shown in Fig. \ref{fig-hr} as a red-thick-solid line.
The referenced 14 M$_\odot$ single stellar model shown by the black-thick-solid line
ends its life as a red supergiant as expected from the canonical stellar evolution theory.
On the other hand, the final position of the merger model
is well within the observational uncertainties of Sk -69$^\circ$202.
The final mass of the model is 18.29 M$_\odot$.
Moreover, the surface chemical compositions before the envelope deflation
are 0.139, 3.47, and 1.21 for He/H, N/C, and N/O, respectively,
which match with observation as well.
When the formed Fe core of the progenitor becomes a neutron star, the rotation period is 74.6 ms which is consistent with the typical order.
Thus, applying our simplified prescription of the slow merger,
we can produce a satisfactory stellar model for the progenitor of SN 1987A.
The best model can explain most observations compared with the previously works quantitatively.
The other lines shown are models with different $M_2$.

\begin{figure}
 \begin{center}
    \includegraphics[angle=-90, scale=0.35]{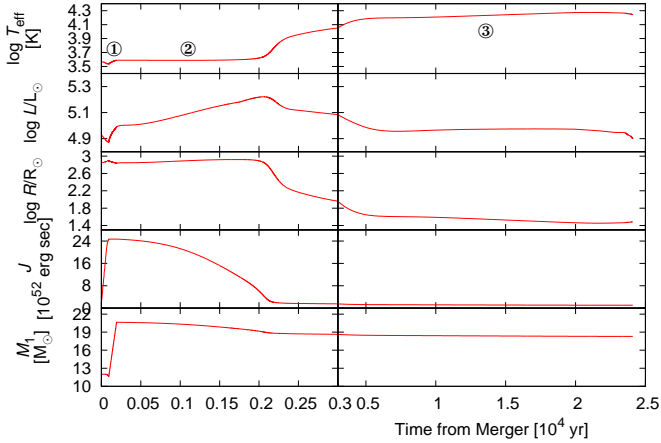}
    \caption{The time evolution of
    the effective temperature,
    the luminosity,
    the radius,
    the spin angular momentum,
    and the mass of the satisfactory model.}
    \label{fig-rot}
  \end{center}
\end{figure}

For the satisfactory model, time evolutions of
the effective temperature, the surface luminosity, the stellar radius,
the total angular momentum, and the mass
are shown in Fig. \ref{fig-rot}.
At first the slow merger takes place ($\textcircled{\scriptsize 1}$).
The hydrogen/helium transition region is significantly modified
due to the chemical mixing and the accretion of the secondary matter.
Since the internal thermal structure cannot adjust to this short-timescale modification,
the thermal equilibrium in the star is strongly violated.

As the slow merger completes, the surface evolution reaches the high luminosity phase
($\textcircled{\scriptsize 2}$).
This lasts $\sim$ 1000 yr, keeping a surface state of
log $T_{\rm eff} \sim 3.6$, log $L$/${\rm L}_{\odot} \sim 5.1$,
and log $R$/${\rm R}_{\odot} \sim 3.0$.
Limitation of the duration is due to the lack of the sustainable energy source in the star.
Internal luminosities at shell helium or hydrogen burning regions
are up to log $L$/${\rm L}_{\odot} \sim 4.9$.
Instead the large surface luminosity is supplied by
the transported thermal energy from the base region of the convective envelope.
Due to the gravity amplification resulting from the mass accretion,
the inflating convective envelope progressively falls onto the inner mantle,
decreasing the entropy and supplying the luminosity.
After the whole part of the convective envelope falls down, the stellar surface shrinks.
This is coincidently observed as increasing of the effective temperature.

The rate of the change becomes slower and slower
as the effective temperature gets closer to $\rm{log}$ $T_{\rm eff} \sim 4.2$.
Finally the star reaches a new thermally equilibrated structure.
The star enters into the quasi-stable radiative phase,
which lasts until the star collapses ($\textcircled{\scriptsize 3}$).

\subsection{Evolution of the spin angular momentum}

Efficient mass ejection results from angular momentum injection to the envelope.
Although the magnitude, $J_{\rm add} = 3 \times 10^{53}$ erg s, for the model shown here
is about one-tenth of a typical initial orbital angular momentum,
the result will be still informative to see
what will happen if the large quantity of the angular momentum is given to the stellar envelope.

First of all, almost all the additional angular momentum is lost during high luminosity phase.
In this phase, a transition from convective to radiative in the envelope occurs.
Then angular momentum is transported near the surface efficiently, and the most is concentrated there.
Thus the large angular momentum added to the envelope can be lost
by ejecting a relatively small amount of the envelope of $\sim$ 2.7 M$_\odot$ (see \citealt{heger+98} for more details of the mechanism).

It is noteworthy that
the rotation speed after the envelope deflation is just $\sim$ 20\% of the critical velocity,
despite the stellar radius significantly decreases by one order of magnitude.
The speed  $\sim 80$ km s$^{-1}$ is less than an observational upper limit on the pre-supernova rotation velocity 100 km s$^{-1}$ estimated by  \citet{parth06}.
As shown before, the base of the convective envelope progressively falls onto
the inner stellar mantle during the high luminosity phase.
This descending matter initially spins up.
However, the angular momentum of the descending material is soon transported to
the still inflating convective envelope due to efficient magnetic viscosity.
As a result, the almost all adding angular momentum,
which is initially distributed over the convective envelope,
is ejected through the inflating surface.
The blue deflated star finally
contains only $\sim 5 \times 10^{51}$ erg s of the angular momentum in its envelope.

\subsection{Results with $J_{\rm add} \geq 10^{54}$ erg s}

The larger the $J_{\rm add}$ is, the larger amount of mass is ejected.
We find it is difficult to make a BSG progenitor in our calculation
even with $J_{\rm add} = 1 \times 10^{54}$ erg s,
which is still about one-third of a fiducial value,
since too large mass of $\sim$ 5 M$_{\odot}$ is lost.
We summarize results of evolution calculations for selected models in Table \ref{tab-model},
in which $M_{\rm fin}$ and $\Delta M$ are the final mass and the mass lost after the merger.
Positions in the HR diagram of three models out of them are compatible with the observation,
however all of them are models with $J_{\rm add} < 1 \times 10^{54}$ erg s.
On the contrary, all models with $J_{\rm add} \geq 1 \times 10^{54}$ erg s finally become RSGs.

\begin{table*}
  \caption{Summary of the results.}                                                     
  \begin{tabular}{ccccccccc}                                                                                                           
  \hline \hline
  \multicolumn{9}{c}{$M_{1}$ = 14 M$_\odot$, $M_{\rm in} = $ 4.6 M$_\odot$,
   log $R_{\rm out}/{\rm R}_\odot = 2.75$}  \\
  \hline                                                                                                                     
   $J_{\rm add}$  & $M_{2}$ & $M_{\rm fin}$ & $\Delta M$ & log $T_{\rm eff}$ &
   log $L/{\rm L}_{\odot}$  & He/H & N/C & N/O   \\
   $[\times 10^{53}$ g cm$^2$ s$^{-1}$] &  [M$_\odot$] & [M$_\odot$] & [M$_\odot$] & [K]  &
    - & - & -  & - \\
  \hline 
   1.0 & 6.0 & 16.87 & 1.14 & 4.173 & 4.861 & 0.145 & 3.56 & 1.23  \\
   1.0 & 7.0 & 17.92 & 1.09 & 4.225 & 4.900 & 0.142 & 3.50 & 1.22  \\          
   1.0 & 8.0 & 18.96 & 1.05 & 4.310 & 4.957 & 0.140 & 3.47 & 1.21  \\                                                                            
   3.0 & 7.0 & 16.07 & 2.94 & 3.640 & 4.381 & 0.145 & 4.29 & 1.29  \\
   3.0 & 8.0 & 17.27 & 2.74 & 4.168 & 4.854 & 0.141 & 3.51 & 1.22  \\          
   3.0$^1$ & 9.0 & 18.29 & 2.72 & 4.244 & 4.902 & 0.139 & 3.47 & 1.21  \\
   3.0 &10.0 & 19.35 & 2.66 & 4.336 & 4.994 & 0.138 & 3.46 & 1.21 \\
   5.0 & 8.0 & 16.09 & 3.92 & 3.632 & 4.423 & 0.144 & 5.01 & 1.34  \\
   5.0 & 9.0 & 17.20 & 3.81 & 4.187 & 4.872 & 0.140 & 3.55 & 1.23  \\          
   5.0$^2$ & 9.5 & 17.69 & 3.82 & 4.257 & 4.915 & 0.139 & 3.52 & 1.23  \\             
   7.0 & 8.0  & 15.43 & 4.58 & 3.577 & 4.946 & 0.147 & 3.83 & 1.29  \\
   7.0 & 9.0  & 16.40 & 4.97 & 4.123 & 4.821 & 0.141 & 3.67 & 1.25  \\          
   7.0$^2$ & 10.0 & 17.41 & 4.60& 4.253  & 4.890 & 0.142 & 3.68 & 1.26  \\             
   10.0 & 10.0 & 16.83 & 5.18 & 3.984 & 4.773 & 0.141 & 3.68 & 1.26  \\          
   20.0 & 10.0 & 15.90 & 6.11 & 3.577 & 4.955 & 0.140 & 3.57 & 1.25  \\          
   30.0 & 10.0 & 15.72 & 6.29 & 3.577 & 4.954 & 0.142 & 3.93 & 1.29  \\                     
  \hline         
  \end{tabular}

  \small{$^1$ The satisfactory model.
  $^2$ Models match with observational constraints except for the final mass.}
  \label{tab-model}
\end{table*}

Our current method to model the critical rotation mass ejection, however,
can be improved into several directions.
Especially, the geometry of mass ejection can be modified.
The rotational mass loss is assumed to take place as an enhancement of
spherical wind mass loss in this work.
If disk-like mass ejection takes place instead,
the ejected mass can possess much higher specific angular momentum than now,
and the total ejecta mass will be reduced.
Therefore, in order to avoid the too much mass ejection we have pointed out,
it is important to consider more realistic geometry of mass ejection
from a critically rotating stellar envelope.

The time scale of dissipation of the additional angular momentum inside the envelope can also be improved.
We assume that the added angular momentum is instantaneously redistributed over the convective envelope.
However, the dissipation will be done through, for example, viscous or tidal dissipations.
If the timescale is longer than the duration of the high luminosity phase, which is about a thermal time,
the additional angular momentum will be locally trapped.
This will affect the mass loss rate.
Besides, mass loss rate from a critically rotating surface is
one of the highly uncertain ingredient in our calculation.
Although we currently set the upper limit using the thermal time (eq.(\ref{og-limit})),
there might be no such a limit for the dynamical mass ejection.
However, a much higher mass loss rate may not affect the total mass lost,
since the mass loss rate enhancement ceases when all of the additional angular momentum is lost.


\section{Conclusion}

The evolutionary properties of Sk -69$^\circ$202, the progenitor of SN 1987A, are investigated.
As the red-to-blue evolution has not occurred in neither single rotating models
nor binary mass accretion models with realistic parameters,
we instead consider the slow merger scenario to include the effects of 
the strong internal mixing and the envelope mass enhancement in the modeling.
In our one-dimensional approach, in addition to these two effects,
the acceleration of the envelope spin for the first time is taken into account.

Without any fine tuning, which is often required to produce the surface evolution
in single evolutionary models,
observational constraints of
the red-to-blue evolution at $2 \times 10^4$ yr before core collapse
and the surface chemical composition and the total mass at the explosion are
well reproduced in our model.
This strongly indicates the robustness of this scenario.
Besides, we have found the mass lost to eject the fiducial amount of angular momentum of
$\sim 3 \times 10^{54}$ erg s is too large to built a blue static envelope in the current modeling.
This suggests that disk-like mass ejection takes place in reality
when a stellar surface reaches the critical.\\

KT was supported by JSPS Overseas Research Fellow.
This work was supported by Grant-in-Aid for Scientific Research on Innovative Areas (No. 26104007)
and in part by the Grant-in-Aid for Scientific Research (No. 26400271).













\bsp	
\label{lastpage}
\end{document}